\documentclass[preprint,12pt]{elsarticle}

\usepackage{amssymb}

\usepackage{lineno}

\journal{Nuclear Instruments and Methods A}

\begin{document}

\begin{frontmatter}



\title{ Behaviour in Magnetic Fields of Fast Conventional and Fine-Mesh 
Photomultipliers.}

\author[1]{M. Bonesini} 
\ead{Maurizio.Bonesini@mib.infn.it, corresponding author} 
\author[1]{R. Bertoni}  
\author[2]{A. de Bari}   
\author[2]{M. Rossella} 
\address[1]{INFN -- Sezione di Milano Bicocca, Dipartimento di Fisica G. Occhialini, Piazza Scienza 3, Milano, Italy}
\address[2]{INFN -- Sezione di Pavia, Dipartimento di Fisica Nucleare e Teorica, via A. Bassi 6, Pavia, Italy}
\begin{abstract}
The performance of  both conventional and fine-mesh
Hamamatsu photomultipliers  has been measured inside  moderate
magnetic fields. This has allowed the
test of  effective shielding solutions for photomultipliers,
to be used in time-of-flight
detectors based on scintillation counters. Both  signal amplitude
reduction or deterioration of the timing properties inside magnetic fields
have been investigated.
\end{abstract}

\begin{keyword}
time-of-flight detectors \sep photomultipliers 


\end{keyword}

\end{frontmatter}


\section{Introduction}
A simple parametrization of the timing resolution of time-of-flight
(TOF) detectors, based on scintillation counters, is given in 
references~\cite{tof,tof1}:
\begin{equation}
\sigma_{t}=\sqrt{\frac{\sigma_{sci}^2 + \sigma_{PMT}^2 + \sigma_{pl}^2}
{N_{pe}} + \sigma_{elec}^{2}}
\label{tof:res}
\end{equation}
where:
$\sigma_{sci}$ is the  scintillator response, 
$\sigma_{PMT}$ the  photomultiplier (PMT) jitter,
$\sigma_{pl}$  the   path 
lengths variations, 
$\sigma_{elec}$  the jitter of the electronic readout system and
$N_{pe}$ is the average number of  photoelectrons.
The dominant factors for $\sigma_t$ are $N_{pe}$ and the counter dimensions
(mainly its length L), responsible for path lengths fluctuations.  
Below 100 ps, contributions such as $\sigma_{PMT}$ 
become increasingly important.

As the PMT signals of a TOF system are commonly fed into a 
time-to-digital converter (TDC)  
after a leading edge 
discriminator, a relevant reduction in PMT gain (as happens inside magnetic fields)
is to be avoided, to
prevent increasing and less reliable time-walk corrections.

As a consequence, the operation inside even moderate magnetic fields 
(up to some hundreds of Gauss), such as stray fields of magnets,  puts severe 
requirements on the properties of the used photomultipliers, for what
concerns  gain and timing properties. 
The more common solution for conventional PMTs is to use  local shieldings,
made  of a cylindrical envelope of high permeability material, 
such as $\mu-$ metal. The shielding factor is proportional to the relative
magnetic permeability $\mu_{r}$ 
of the used material multiplied by some (non-negligible) geometric factor.
High permeability alloys are available, but saturation effects limit their
use to low external magnetic fields (up to 50-100 Gauss). To reduce this
effect the shield radius may be increased. As these materials 
are quite expensive,
a common solution is to use in addition to a thin, high permeability inner 
shield, a thick soft iron external shield. These composite shieldings are
quite effective 
to suppress the orthogonal component of the magnetic field $\bf{B}_{\perp}$,
but problems may arise for the axial component $\bf{B_{\parallel}}$, 
parallel to the PMT axis.
While formulas to compute the ``shielding factor'' for orthogonal 
fields are easily 
available \cite{pmt-handbook}, results for the shielding of the axial 
component of a magnetic field are less common. 
 
Fine-mesh dynodes PMTs are instead nearly insensitive to 
the axial component
of the magnetic field $\bf{B_{\parallel}}$, 
but may have problems with an increasing transverse component $\bf{B}_{\perp}$.
These tubes become totally unusable for orientations of the PMT axis, 
with respect
to the magnetic field, bigger than a critical value $\theta_c$ 
($\sim 45-60^0$) even in small magnetic fields ( $\sim 150-200$ Gauss).
The structure of these PMTs is based on a sequence of  fine-mesh electrodes,
where the incoming electrons are multiplied by secondary emission. 
The electrons in the final
stage are  collected by the anode as output signal. 
The distance between the first dynode and the photocathode is $\sim 4$ mm,
while the distance between successive dynodes is $\sim 1$ mm.

In the following, results on conventional 1" R4998 and R9800 PMTs 
and fine mesh
1", 1.5", 2" R5505, R7761 and R5924 PMTs from Hamamatsu will be shown
(see table \ref{tab1} for their main properties).

They have been of interest in detector optimization for the timing 
counter~\cite{meg} of the MEG experiment at PSI~\cite{meg1} and the TOF 
detectors~\cite{bertoni,bonesini1} of the MICE 
experiment at RAL~\cite{mice}, devised to study ionization cooling for the proposed
Neutrino Factory \cite {bonesini3} and Muon Collider~\cite{neuffer}. 
\begin{table*}
{\footnotesize
\begin{center}
\begin{tabular}{|c|c|c|c|c|c|}
\hline
              & R4998  & R9800& R5505 & R7761     & R5924         \\ \hline
PMT type      &conventional& conventional & fine-mesh & fine-mesh & fine-mesh \\
Tube diameter &1"      & 1"    & 1"    & 1.5"       & 2"             \\
No. stages    & 10     & 8 & 15    & 19         & 19             \\
Q.E. at peak  & .20    &.25   & .23   & .23        & .22            \\
Gain (B=0 T) typ. & $5.7 \times 10^6$  & 
$1.0 \times 10^6$ & $5 \times 10^5$ & $1.0 \times 10^7$  & $1.0 \times 10^7$ \\
transit time (ns)& 10  & 11 & 5.6 &7.5 & 9.5  \\
Risetime (ns) & 0.7  & 1.0 & 1.5            &    2.1              &  2.5    \\
TTS (ns)      & 0.16 & 0.27 & 0.35           &    0.35             &  0.44     \\  
HV (V) (max value)   & -2500 & -1500 & +2300 & +2300 & +2300 \\
\hline
\end{tabular}
\caption{Main properties of the Hamamatsu PMTs under test. Conventional PMTs
(R4998 and R9800) have a linear-focussed structure for dynodes.}
\label{tab1}
\end{center}}
\end{table*}

All PMTs were delivered as assemblies with a passive or active divider base 
and a 1 mm-thick  $\mu$-metal shield.

For PMT timing the relevant properties are the transit time spread (TTS) and the 
PMT  size that may have an effect on it, by an increasing photocathode area 
(with a wider photoelectrons spatial distribution). 
For this reason this paper will report mainly about the smaller size (1")
PMTs, that have better timing properties. 

\section{Test results for conventional PMTS}

Systematic studies of conventional PMTs have been performed, using 
a built on purpose solenoid of 23 cm inner diameter,  
40 cm length~\footnote{
TBM srl, Uboldo (VA), Italy}. 
The solenoid being a resistive magnet, 
special care was put into the thermal resistance of the assembly
(up to $100^{0} \ C$), using special insulating paints. 
For part of the tests a Digimess 3040 laboratory 
power supply (0-32 V, 0-40 A) was used. For larger field amplitudes
an Eutron power supply (0-32 V, 0-100 A) was employed.   
The main constraint was the heating of the windings that limited
the maximum circulating current to about 55 A (corresponding to
a maximum field of $\sim $ 600 Gauss), due to the
increase of conductor resistance giving higher voltage
drops (up to the maximum allowed value of $\sim 30 \ V$).
The large open bore allows tests 
both with field lines orthogonal or parallel to the
PMT axis. 

\begin{figure*}[hbt]
\begin{center}
\includegraphics[width=\linewidth]{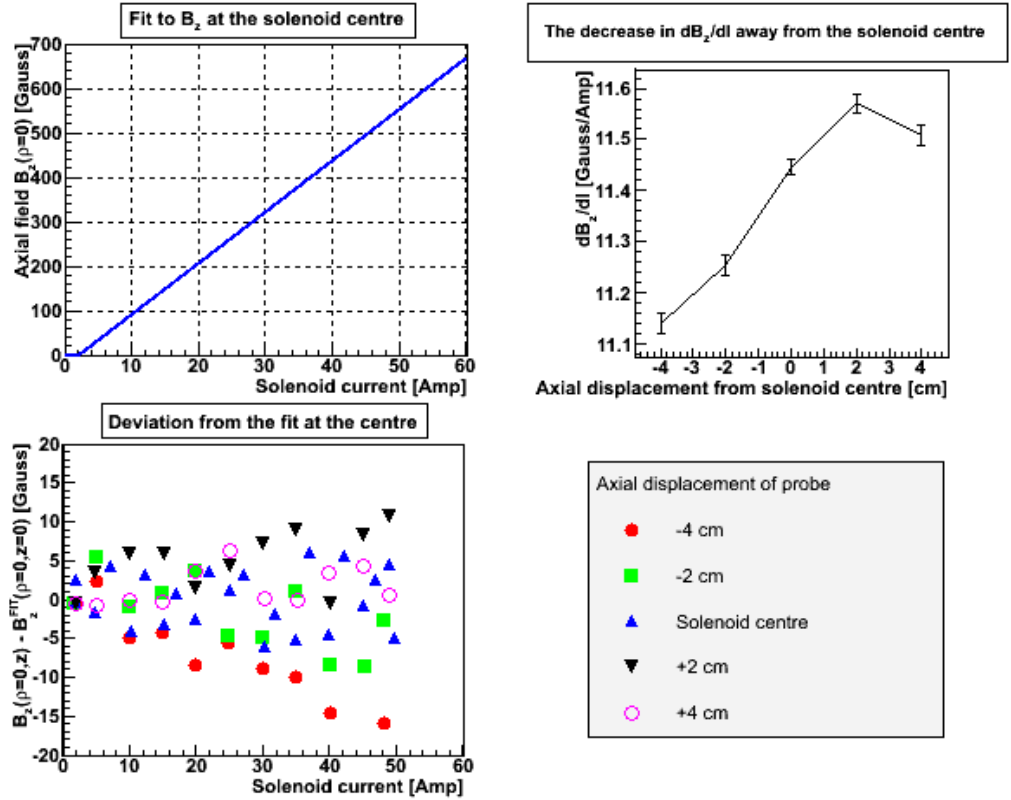}
\caption{ Upper panel: (left) calibration of the longitudinal field at the geometrical center of the test solenoid, as function of
the applied current; (right) variation of the calibration slope as function 
of the axial displacement respect to the solenoid center. Lower panel: 
difference of the magnitude of the {\bf B} field as respect to $z=0$ (solenoid
center, z axis along the solenoid) as a function of the circulating current.}
\label{fig:mag2}
\end{center}
\end{figure*}

The magnetic field was measured with a gaussmeter~\footnote{
Hirst GM04 model, with transverse Hall probe},
with a relative accuracy better than $1 \%$. The {\bf B} field calibration curve 
is shown
in figure \ref{fig:mag2} at the center of the solenoid and at different
$z$ positions shifted along its longitudinal axis.  
From simulations using a finite length solenoid approximation and 
performed measurements,
a field uniformity better than $3 \%$ may be assumed  
at the center of the test solenoid in a volume of about
$5 \times 5 \times 10 \ cm^3$. 

A fast light pulse ($\sim 1$ ns width, to simulate a typical scintillator
signal)~\footnote{ 
The system 
is based on a Nichia NDHV310APC violet laser diode, driven by 
an AvtechPulse fast
pulser (type AVO-9A-C, with $\sim200$ ps
risetime) and an AVX-S1 output module. 
Laser pulses at $\sim 409 \ nm$, 
with a FWHM between $\sim 120$ ps and $\sim 3$ ns were obtained.}
was sent to the centre of the 
PMT photocathode via a 3 m long multimode 3M TECS FT-110-LMT optical
fiber~\footnote{with a measured dispersion of $ \leq 15$ ps/m, 
see \cite{bonesini}.} inserted in a small Plexiglas cover 
in front of the PMT window.
To provide light signals of various intensities, the laser spot was injected 
into the optical fiber 
by a 10x Newport microscope objective, after  removable absorptive 
neutral density filters.
A broadband beamsplitter (BS) divided the laser beam to give $ 50 \% $
of light on the fiber injection system and $ 50 \%$ on a 
fast Thorlabs DET10A photodiode (risetime $\sim 1$ ns), 
to monitor the laser system stability.
Tests were carried out  with a signal corresponding to about 150-300 
photoelectrons \footnote{The number of photoelectrons ($N_{pe}$) was 
estimated via absolute gain measurement.
This number was cross-checked with  powermeter 
measurements.}: a
typical value for a minimum ionizing particle (MIP) crossing a 1" to 2"-thick
scintillator.
The optical power was periodically monitored with an 
OPHIR NOVA laser powermeter.

\begin{figure*}[hbt]
\begin{center}
\includegraphics[width=0.8\linewidth]{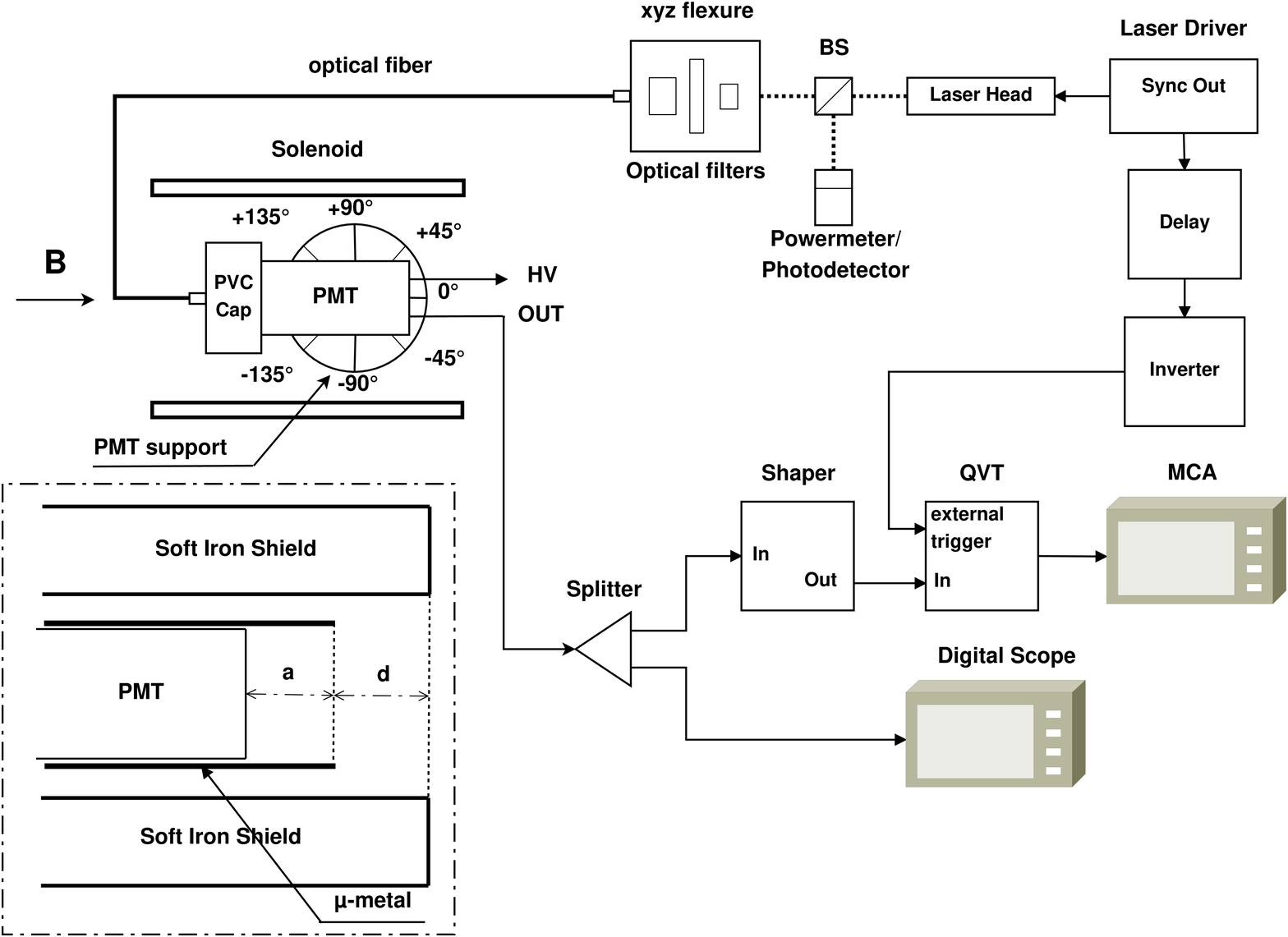}
\caption{ Layout of the test setup for PMT measurements (not in scale). 
In the inset a schematic of the PMT shielding is shown. The distance $a$ 
is typically 3 cm for the conventional PMTs under test and the distance
$d$ is between 0 and 3 cm.}
\label{fig:setup}
\end{center}
\end{figure*}

The experimental setup is shown in figure \ref{fig:setup}.
The PMT signal was input after a passive splitter to 
a digital scope~\footnote{ Tektronix TDS 754C with 500 MHz bandwidth, 
2 Gs/s sampling rate or Tektronix DPO7254 with 2.5 GHz bandwith, 
40 Gs/s sampling rate.}
triggered by the
laser  synchronization  output signal (sync. out) with a maximum jitter 
of $ \pm 15$ ps with respect 
to the delivered optical pulse 
 and  to a Silena 8950 Multichannel Analyzer (MCA).
For timing measurements, the MCA chain 
was used with a Silena 7422 charge-amplitude-time converter (QVT).
The PMT anode signal after a leading edge PLS 707 discriminator provided the
STOP  signal ($t_{STOP}$), while the START signal ($t_{START}$) was 
given by the 
sync out of the laser pulser after a suitable delay and an ORTEC  
pulse inverter.
In timing measurements, the time difference
$\Delta t= t_{START}-t_{STOP}$ is measured. This accounts for delay in cables 
and electronics and jitter in the transit time in the tested PMTs. 
No variation of
this quantity or no deterioration in the FWHM of its distribution, 
after increasing the magnetic field intensity, demonstrates the effectiveness
of the adopted shielding solution.

\subsection{Experimental results}
As conventional photomultipliers may work without major problems inside 
a residual magnetic field of a few Gauss, a preliminary estimate 
of their expected behaviour may be 
obtained from magnetostatics simulations.

\begin{figure*}[hbt]
\begin{center}
\includegraphics[width=0.49\linewidth]{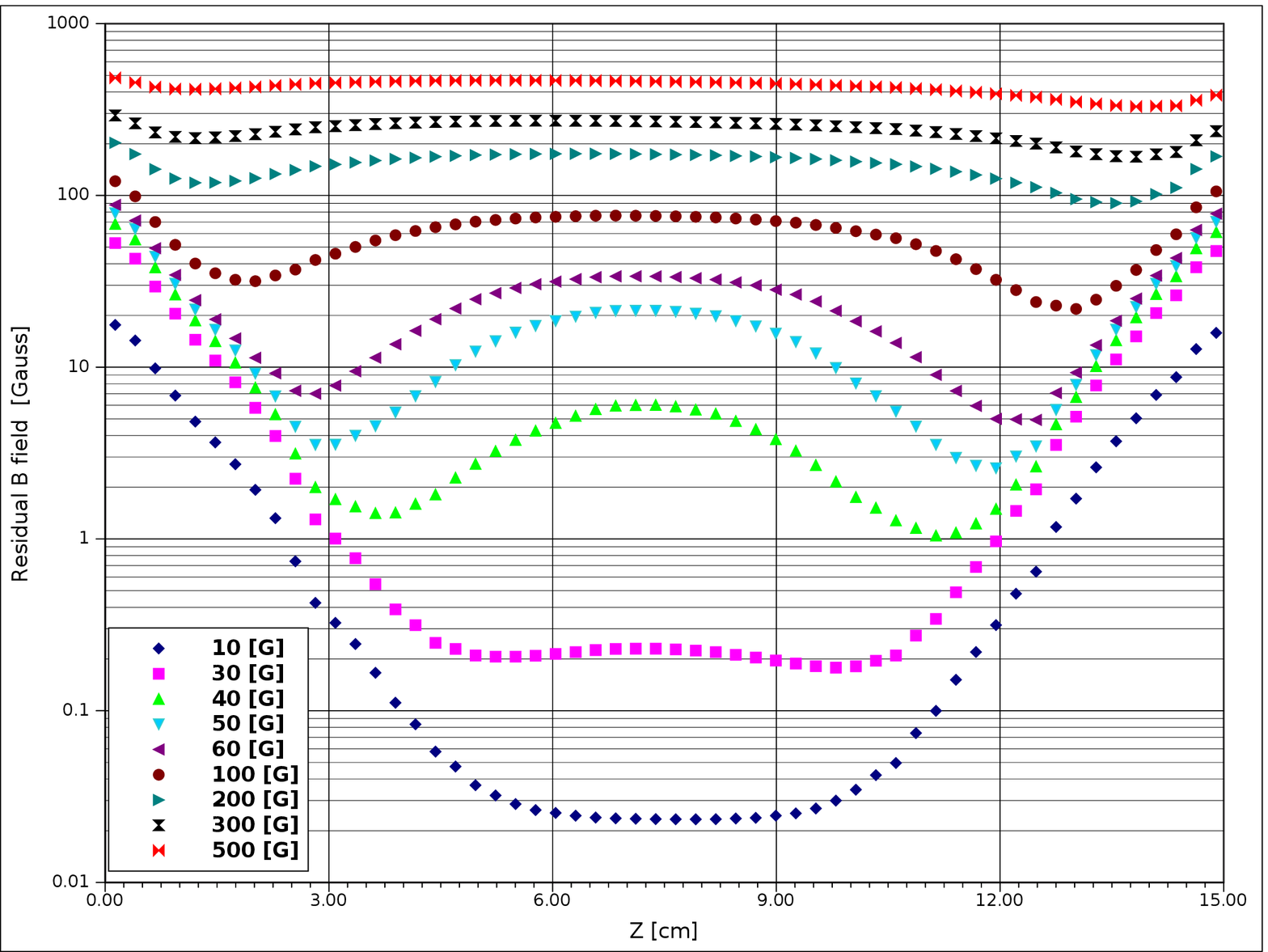}
\includegraphics[width=0.49\linewidth]{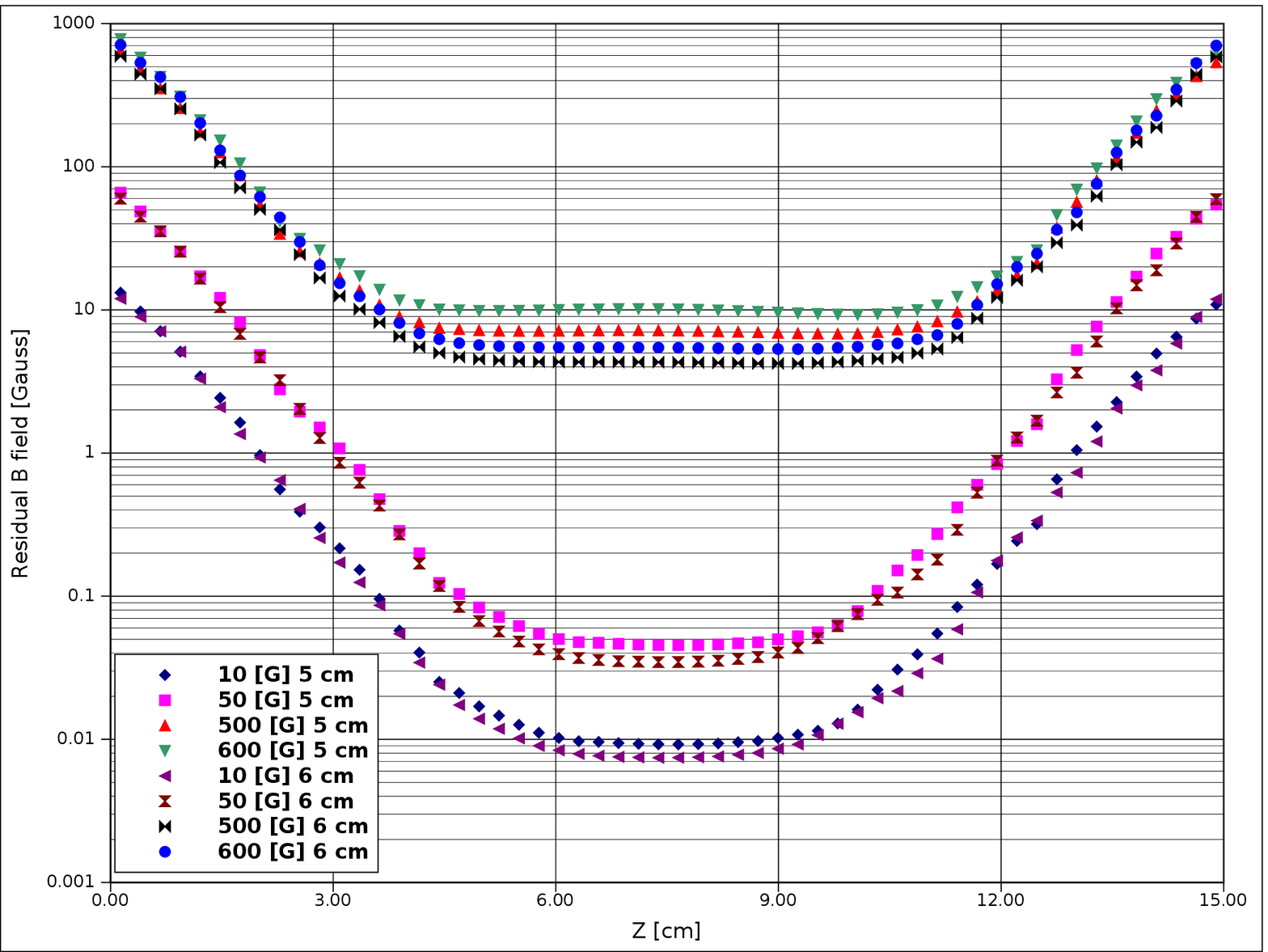}
\caption{ Residual axial  magnetic field, as computed for a simple
1 mm-thick  mu-metal shielding 15 cm long (left panel) or a 15 cm long 
1 mm-thick  mu-metal shielding 
with an external soft iron (AISI 1020 Steel, with maximum 0.23 \% content of 
Carbon) cylinder of radius 5 or 6 cm.}
\label{fig-mag}
\end{center}
\end{figure*}

The residual longitudinal magnetic field inside a $\mu$-metal shielding 
or a composite
shielding made of a soft iron cylinder surrounding a mu-metal shield
may be easily calculated with 2-D magnetostatics simulation programs. 
Results are reported in figure 
\ref{fig-mag} as an example, using the public domain 2-D magnetostatics 
program FEMM 4.2 \cite{simul-mag}.
From the reported simulation results, it is evident that the more the PMT 
is moved inside the external shielding,  the smaller is the value of the
sensed residual magnetic field up to a plateau. This reduction, for the 
$\mu-$metal shielding  works only up to fields $\sim 60$ Gauss, where
the residual magnetic field begins to increase again. 
The situation is more difficult to handle for composite shieldings with 
no azymuthal symmetry, such as the ones with box-shaped soft iron 
external shieldings (discussed later), where more complicate 3-D magnetostatics 
programs may be needed.

In these cases we prefer to rely on experimental 
results, such as the ones reported in the following.
 The PMTs under test were put in the central region of the  
solenoid, using a support
 to incline them  from $0^0$ to  $\pm 90^0$ with respect to the  magnet 
field lines (from {\bf B$_{\parallel}$} to {\bf B$_{\perp}$}).
 Environment light was
accurately masked, to reduce noise. 
Measurements were performed to see gain reduction and possible deterioration in
timing resolution as a function of the magnetic field value ({\bf B}) and the 
relative orientation angle ($\theta$),
between the PMT axis and the magnetic field.
The tested shielding solutions were the $\mu$-metal shielding
only option (1 mm thick $\mu$-metal) and various 
options with additional shields made of soft iron~\footnote{
with typical relative permeability $\mu_{r}^{max} \sim 2000 - 5000$ and
a  reduced area hysteresis loop}  in box or cylindrical shapes.
As the magnetic shielding is mainly a ``mass effect'' we may expect,
for similar transverse dimensions,  
box-shaped shieldings to be more effective than cylindrical ones, having 
more shielding mass~\footnote{This idea was pionereed by the the D0 
experiment in reference \cite{ref:d0}.}.
The tests have been carried out with various shielding configurations:
\begin{itemize}
\item{} only $\mu$-metal~\footnote{with a maximum relative permeability
$\mu^{max}_{r} \sim 200000$.} shielding (1 mm thick, 15 cm long: extending 
3 cm beyond the photocathode area);
\item{} a 15 cm long iron cylinder of diameter 5 or 6 cm with a central hole
        of 3.2 cm diameter (to accomodate inside the PMT assembly with a 
        1 mm thick $\mu-$metal shielding) made of Fe 360
         soft iron~\footnote{S235JR unalloyed steel for magnetic applications,
with a maximal carbon 
content of $0.25 \%$ and a maximum relative permeability 
$\mu^{max}_r \sim 2000$.};
\item{} a 15 cm long iron box of transverse area $ 5 \times 5 \ cm^2$ or
        $6 \times 6 \ cm^2$ with a central hole of 3.2 cm diameter (to
        accomodate the PMT assembly) 
        made of different iron types: Fe360 soft iron
 or Armco.~\footnote{
ARMCO from AkSteel is a pure 
iron with maximal carbon content of $0.025 \%$
and a magnetic saturation ({\bf{J=B-H}}) of 2.15 Wb/$m^2$, much higher than
available commercial soft iron. Its maximum relative permeability is
$\mu^{max}_r \sim 5000$.} 
\item{} the previous two configurations, moving  the PMT assembly 1, 2 or 
3 cm inside the edge
        of the iron shielding
\end{itemize}

Results with only the $\mu$-metal shielding were reported in reference
\cite{bertoni} and show  that R4998 PMTs perform satisfactorily for 
 longitudinal {\bf B$_{\parallel}$} fields up to $\sim 60$ Gauss and 
for orthogonal {\bf B$_{\perp}$}
fields up to $\sim 150$ Gauss.

Inside an orthogonal magnetic field {\bf B$_{\perp}$}, no effect 
larger than a few per-cent is seen
for fields up to $\sim$ 500-600 Gauss using any of the previously 
described composite shieldings.

So we focussed our efforts only on the study of
PMTs inside longitudinal magnetic fields. 

\begin{figure*}[hbt]
\begin{center}
\includegraphics[width=0.40\linewidth]{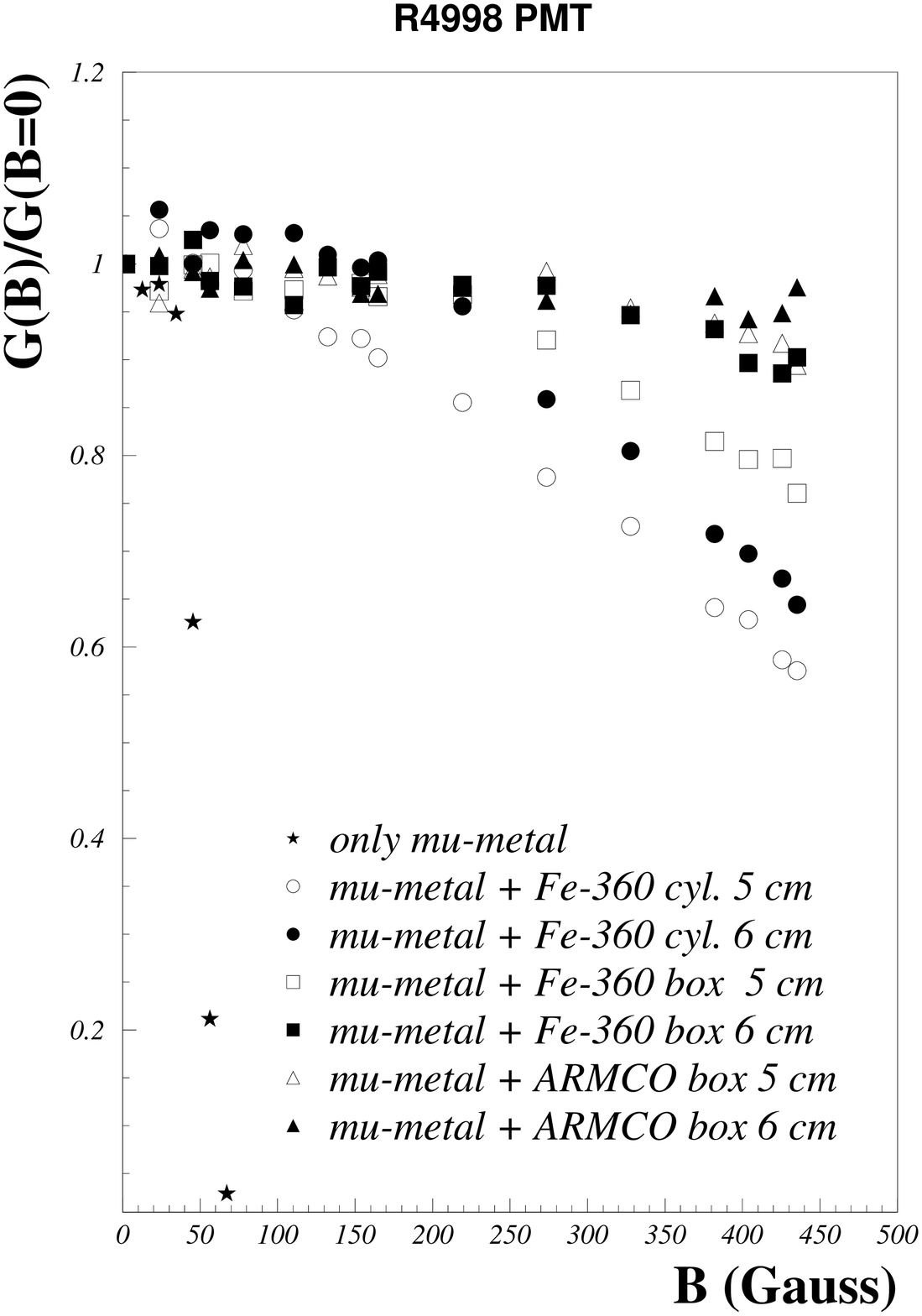}
\includegraphics[width=0.40\linewidth]{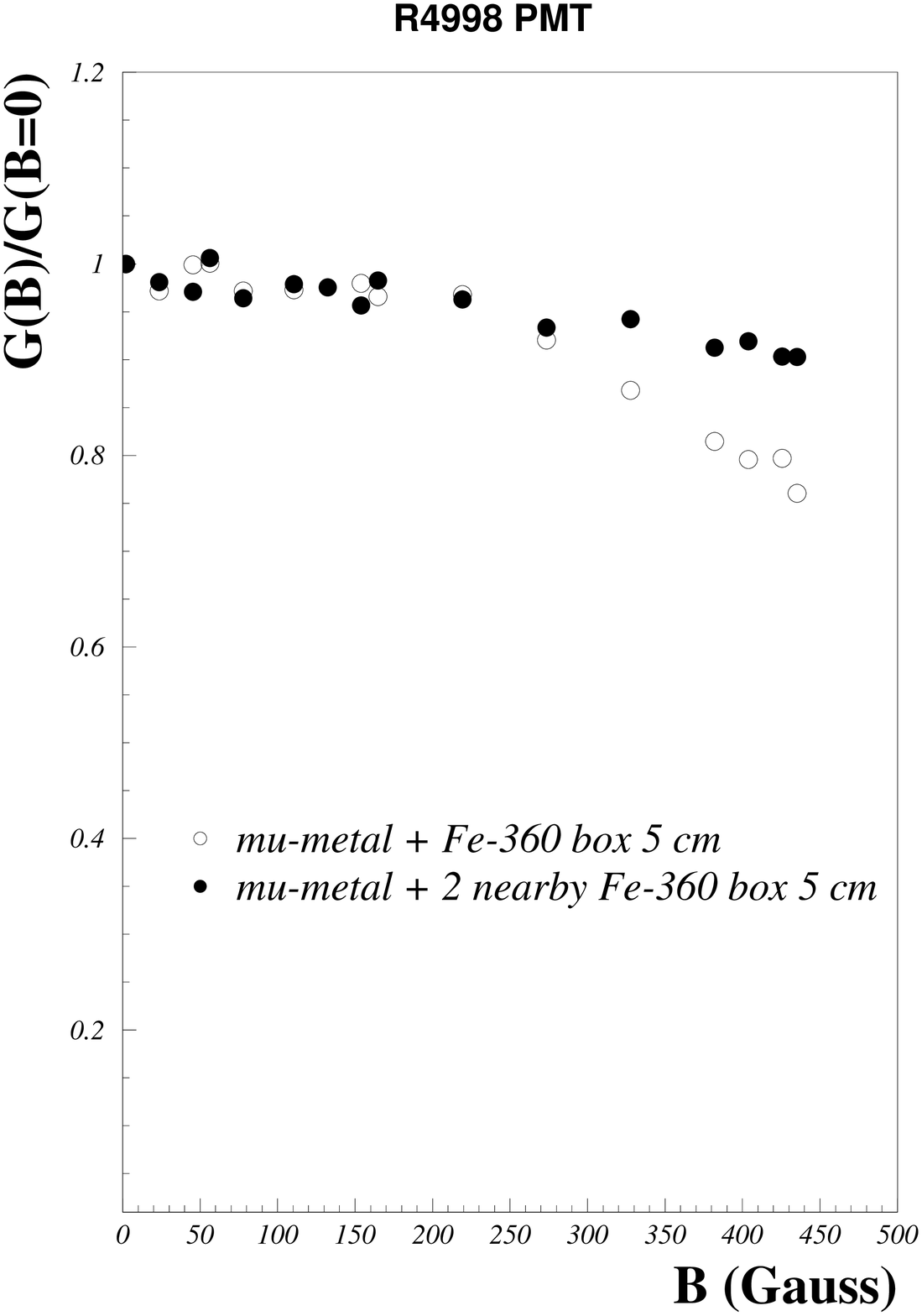}
\includegraphics[width=0.40\linewidth]{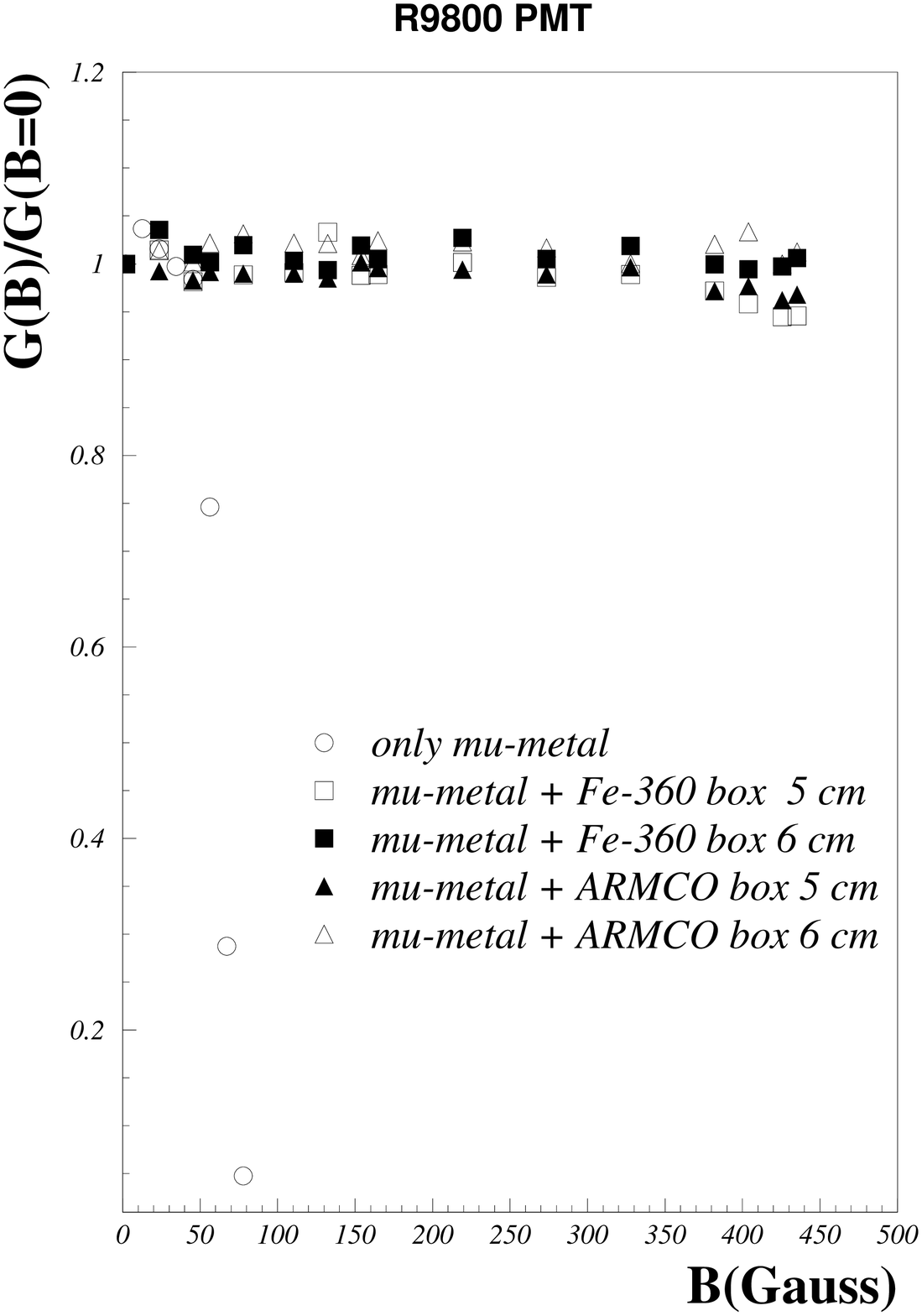}
\caption{ Signal ratio at field B and B=0 G for a  typical R4998 PMT
(top left panel) or a typical R9800 PMT (bottom panel), with
different composite shieldings. The effect of a nearby similar shielding is shown
for a R4998 PMT in the top right panel. 
The B field is along the PMTs axis. The HV was set to
- 2250 (-1350) V.}
\label{fig:shield}
\end{center}
\end{figure*}

Figure \ref{fig:shield} (top left-hand panel) 
shows the signal reduction for a typical R4998 PMT,
with different shielding options: $\mu$-metal only or $\mu$-metal 
with additional
soft iron shieldings, in a longitudinal magnetic field  {\bf B$_{\parallel}$}. 
For equivalent shield sizes, the box-shaped 
iron shielding is clearly more effective than the conventional cylindrical
one. 
 
\begin{figure*}[!hbt]
\begin{center}
\includegraphics[width=0.7\linewidth]{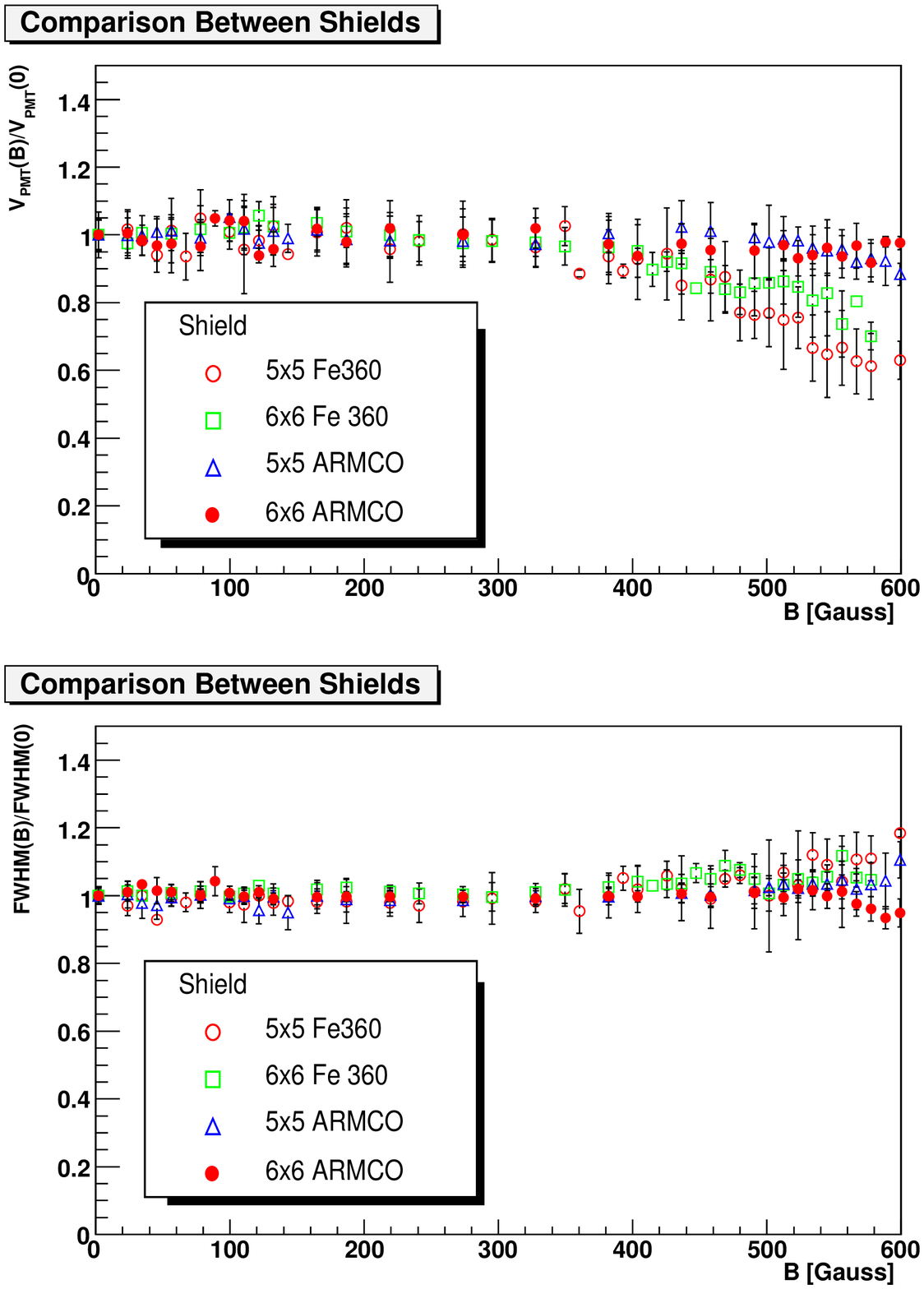}
\caption{ Signal ratio at field B and B=0 G and FWHM ratio at field B and
B=0 G for the timing difference, measured as $\Delta t=t_{START}-t_{STOP}$ with
different  shieldings. The used composite shieldings are:  
a 5 x 5 cm$^2$ box-shaped Fe360 shield ($\circ$), a 6 x 6 cm$^2$ box-shaped 
Fe360 shield ($\Box$),
a 5 x 5 cm$^2$ box-shaped ARMCO shield ($\triangle$) and a 6 x 6 cm$^2$ 
box-shaped ARMCO shield ($\bullet$)
in addition to the inner 1 mm thick $\mu-$metal shielding.
The B field is along the PMTs axis. The plots show the average and rms for a
sample of ten R4998 PMTs. The box shielding does not extend beyond 
the $\mu-$metal (d=0 cm extension).}
\label{fig:test8}
\end{center}
\end{figure*}

\begin{figure*}[hbt]
\begin{center}
\includegraphics[width=0.70\linewidth]{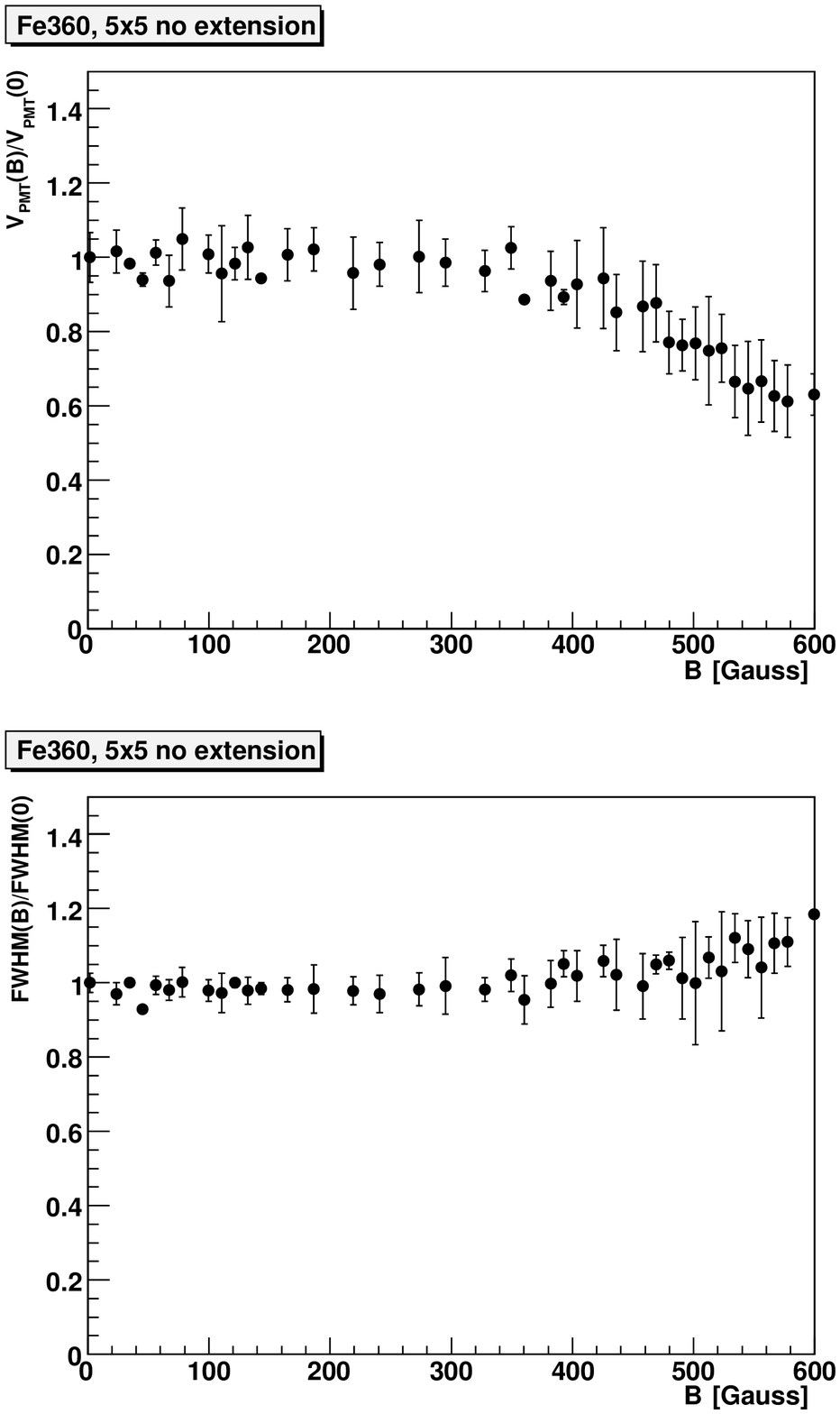}
\caption{ Signal ratio at field B and B=0 Gauss and FWHM ratio at field B and
B=0 Gauss for the timing difference, measured as 
$\Delta t=t_{START}-t_{STOP}$, 
where the 5x5 cm iron box-shielding  extends d=0 cm 
beyond the
end of the mu-metal shielding. The magnetic field {\bf B} 
is along the PMTs axis. 
The plots show the average and rms for a sample of ten R4998 PMTs.}
\label{fig:test3}
\end{center}
\end{figure*}

\begin{figure*}[hbt]
\begin{center}
\includegraphics[width=0.70\linewidth]{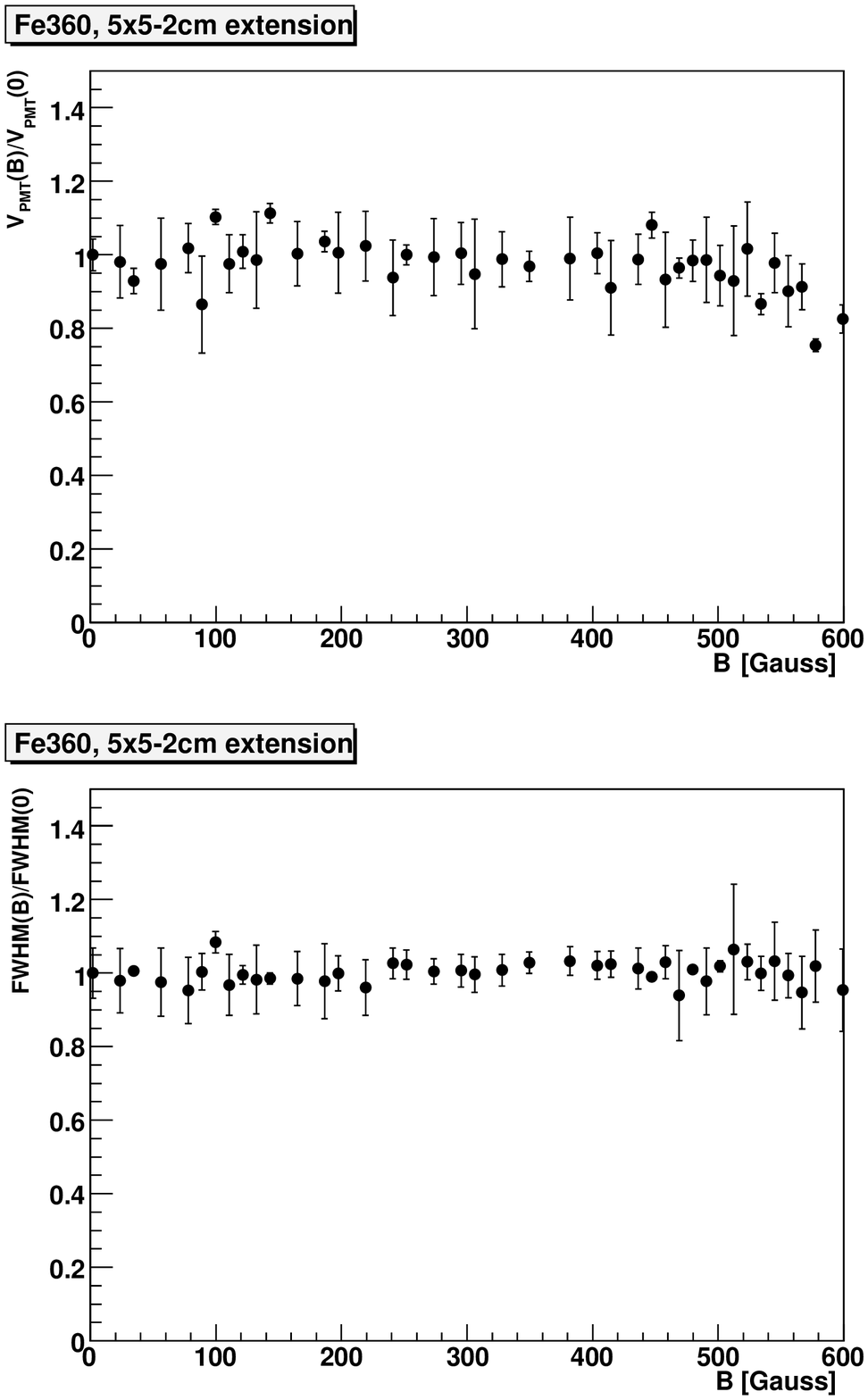}
\caption{ Signal ratio at field B and B=0 Gauss and FWHM ratio at field B and
B=0 Gauss for the timing difference, measured as 
$\Delta t=t_{START}-t_{STOP}$,
where the 5x5 cm iron box-shielding  extends d=2 cm 
beyond the
end of the mu-metal shielding. The magnetic field {\bf B} 
is along the PMTs axis. 
The plots show the average and rms for a sample of ten R4998 PMTs.}
\label{fig:test4}
\end{center}
\end{figure*}

Figure \ref{fig:test8} shows a comparison
for the different box-shaped composite shieldings (signal reduction and 
timing versus the
magnetic field intensity {\bf B}) for the average and rms of a sample
of ten R4998 PMTs. 
The shielding effect may be increased, by moving the PMT assembly more inside
the shielding box, as shown in figures \ref{fig:test3}-\ref{fig:test4} for a 5x 5 cm Fe360
shielding, in addition to the $\mu-$metal one. 

As a general conclusion, we see that:
\begin{itemize}
\item{} very low carbon content iron (ARMCO) is more effective than standard
        Fe360, even if this latter is acceptable for fields up to 500 G 
\item{} box-shaped soft iron shields are more effective than round-shaped
        ones of similar size
\item{} as expected, more massive shieldings (from $5 \times 5 \ cm^2$ to 
        $6 \times 6 \ cm^2$ transverse area) are better
\item{} extending the external soft iron shielding beyond the $\mu$-metal 
        shielding improves the performance
\end{itemize}

The effectiveness of box-shaped iron shielding, in addition to $\mu$-metal, 
for conventional PMTs has been tested also with other types of PMTs. 
The bottom  panel of figure \ref{fig:shield}
shows the results for a typical Hamamatsu 1" R9800 conventional PMT, 
using a $\mu$-metal
shielding only and various types of soft iron (Fe360 or ARMCO) box shieldings.
No deterioration in gain is seen up to $\sim 500$ Gauss 
for {\bf B$_{\parallel}$}~\footnote{R9800 PMTs may have a better behaviour
with respect to R4998 PMTs, due their smaller size and the fewer number of
stages, 8 against 10}.

As long as the PMT signal amplitude 
has a sizeable pulse-height no deterioration in timing
is seen.~\footnote{only with the 1 mm thick $\mu$-metal shielding or the 
1 mm thick $\mu-metal$ shielding + $5 \times 5 \ cm^2$ Fe360 box shielding 
some effect 
was evident in timing ($\Delta t$ or its FWHM), at the $\sim 10\%$ level,
 when the signal amplitude
experienced a reduction of a factor $\sim 10$ in one case 
or two in the other.}        
As local shielding is mainly a mass effect we may expect an improvement if
shieldings of individual PMTs are put in magnetic contact. This is clearly
shown in figure \ref{fig:shield}  (top right-hand panel) 
where an additional iron box of similar 
shape is put 
in magnetic contact with the iron box shielding the PMT.
This is the more common situation for detectors with an array of similar 
channels, where  shielding  may be improved by simply putting all
soft iron individual external shields in magnetic contact.
 
The systematic uncertainties in our  studies are coming mainly from:
\begin{enumerate}
\item{} uniformity of the magnetic field ( better than 3\%),
\item{} stability of the laser pulse intensity (better than 2\%),
\item{} error in positioning of PMTs inside the magnetic field.
\end{enumerate}
The simple mechanics of the system allowed a reproducibility of the 
positioning of the different PMTs' at the level of some mm. 
Every single measurement referred to
about 500-1000 events, giving a negligible statistical error as compared to 
systematics. 
From all the previous sources of errors, we may conservatively estimate a
measurement error  around $ 5-6 \%$. 

No detailed study was done to assess possible azimuthal angle effects 
on the PMTs and hysteresis effects in the shielding.

As expected PMTs behave well for orientation of the {\bf B} field orthogonal 
to the PMT axis ($90^0$), where the shielding effect is maximal, while
along the PMT axis ($0^0$) the gain reduction is more marked. 

Conventional PMTs may be shielded effectively up to longitudinal fields
$\sim 500-600$ Gauss, with simple composite shieldings using $\mu-$metal and
soft iron box-shaped shields. 
For larger magnetic fields, fine-mesh PMTs may be
a better option, even if some additional care must be taken for what
regards their orientation with respect to the magnetic field (see next section for typical results).

\section{Tests on fine-mesh PMTs}
Systematic studies have been carried out also for 1" (R5505), 1.5" (R7761)
 and 2" (R5924) Hamamatsu fine-mesh PMTs (see table \ref{tab1}
for more details). 
They were delivered by Hamamatsu as assemblies with a resistive divider
base and a 1 mm thick $\mu-$metal
shielding,  cut at the photocathode edge. 
A refurbished resistive dipole magnet at LASA (INFN Milano),
with magnetic fields up to 1.2 T and an open gap of 12 cm, was used.
The  test setup was similar to the one employed for the tests
of conventional PMTs, aside the use of a PLP-10 Hamamatsu laser~\footnote{
Laser pulses at $\sim  405 $ nm, 
with 60 ps FWHM pulse width and max repetition rate 100 MHz}.
Data were acquired, after a passive splitter,
both for amplitude measurements (via a CAEN VME V465 QADC) 
and for timing measurements (via a CAEN VME V480 TDC, with a resolution
of 25 ps/ch). The laser sync out, that had a maximum jitter of $\pm 10$ ps
with respect to the optical pulse,  provided in both cases the START/gate signal. 
Tests were usually done with a signal corresponding to about 300
photoelectrons (p.e.).
The optical power was periodically monitored with an OPHIR PD-2A laser powermeter.
The PMTs under test were inserted in the central region of the test magnet, where
the field had a uniformity of $\sim 1 \%$, using plexiglass supports to incline
them up to $60^0$ with respect to the B field lines. 

\begin{figure*}[hbt]
\begin{center}
\includegraphics[width=0.42\linewidth]{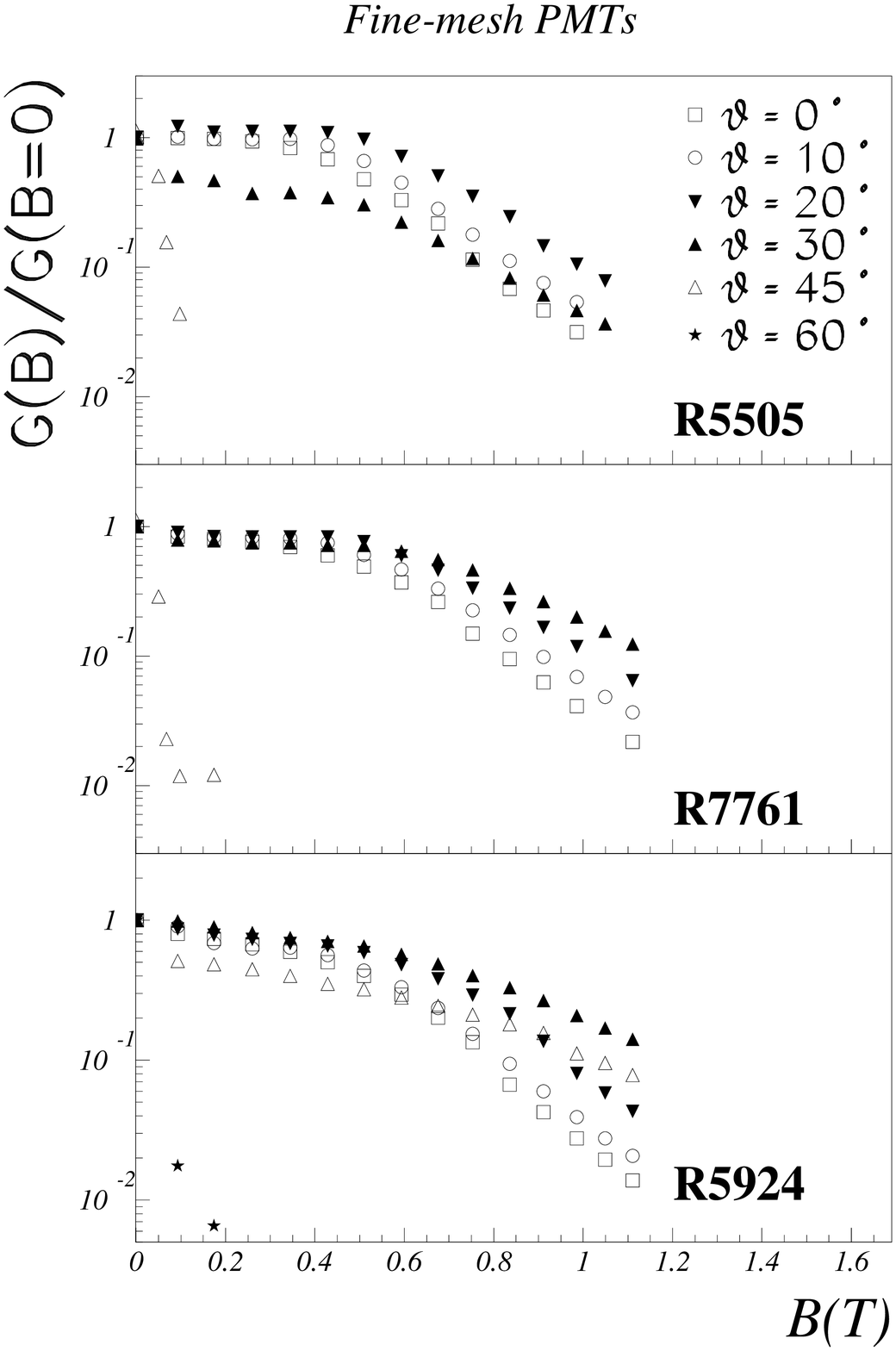}
\includegraphics[width=0.42\linewidth]{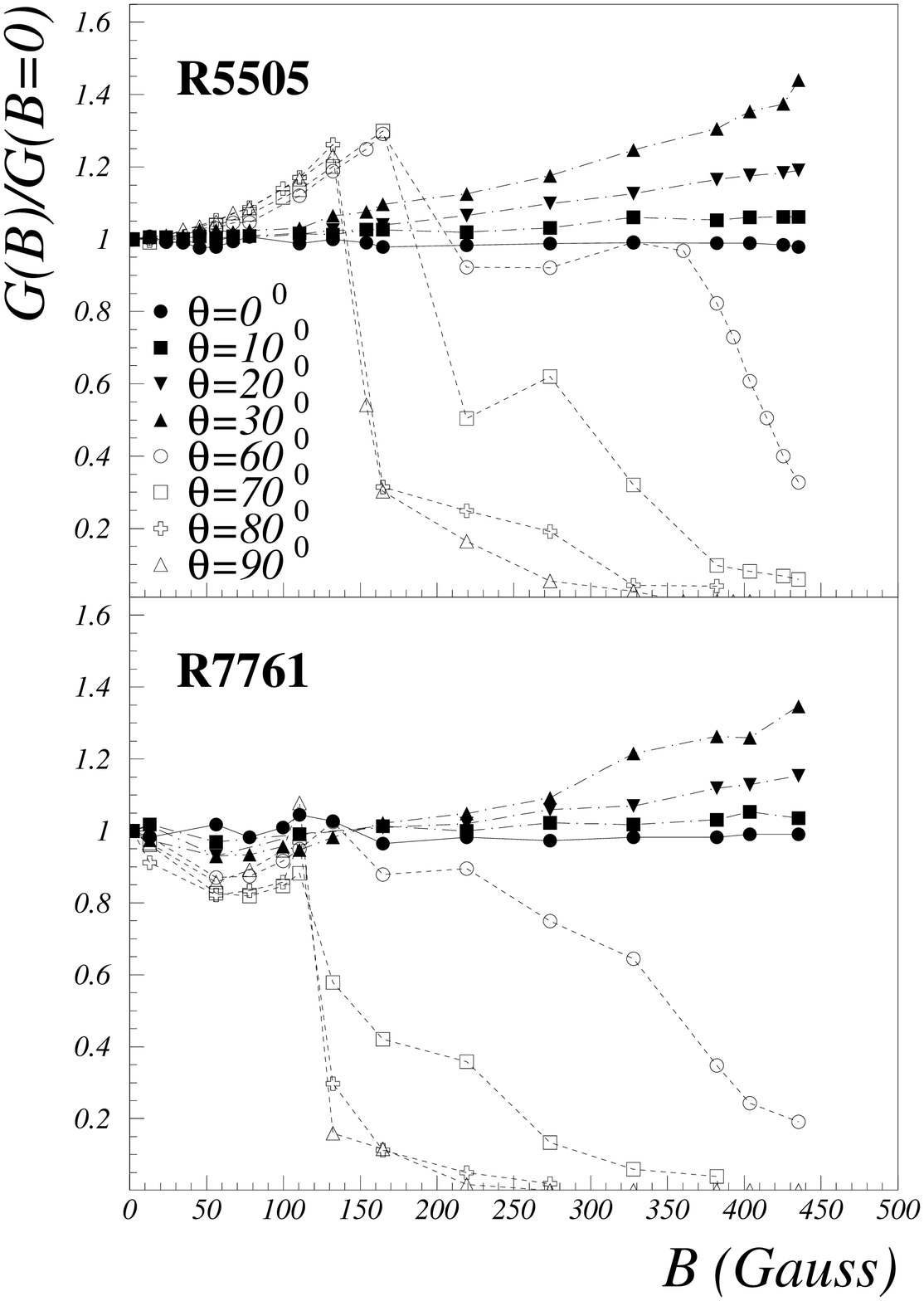}
\includegraphics[width=0.42\linewidth]{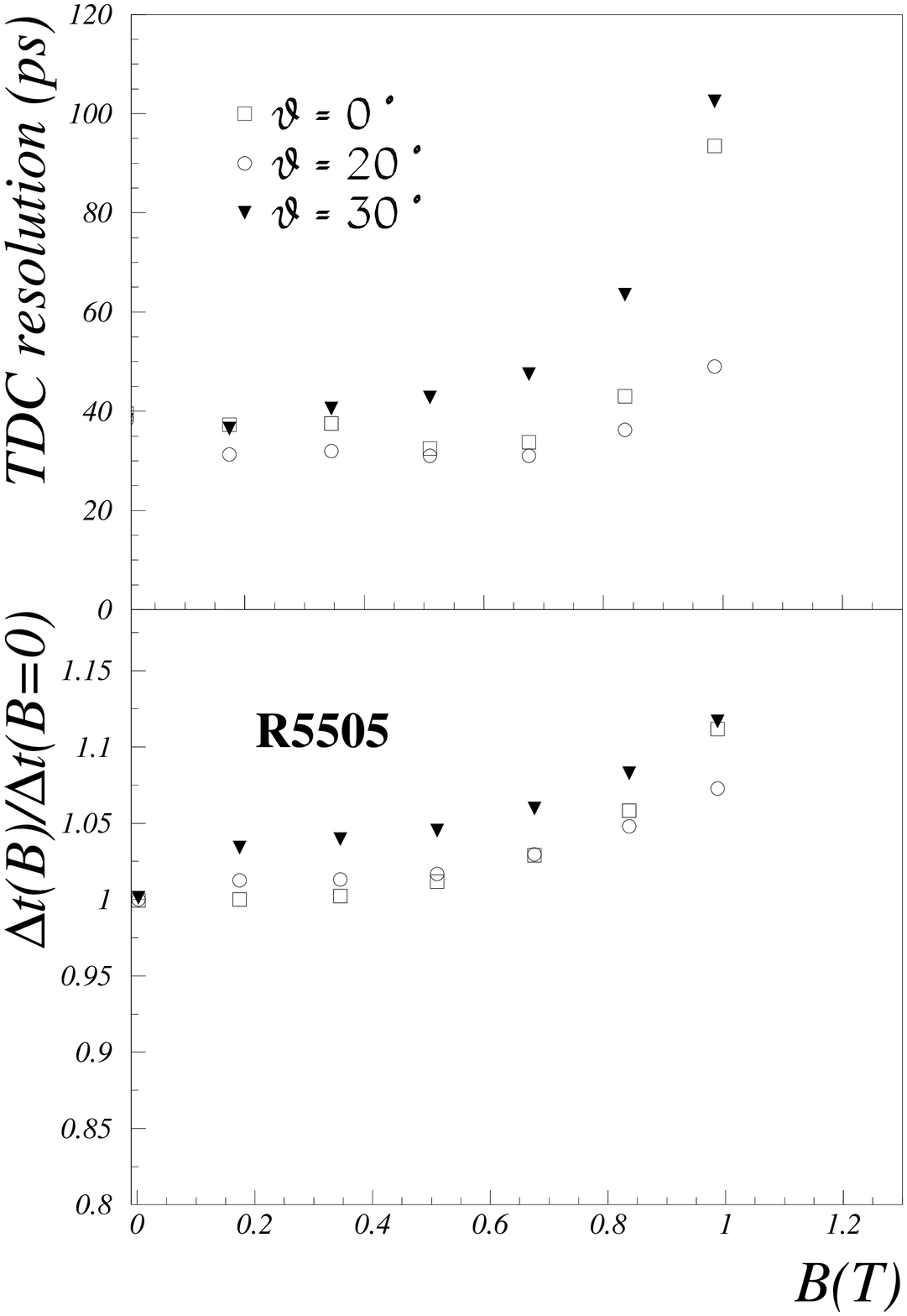}
\caption{Top left panel: behaviour of the gain variation in a high magnetic 
field  
for  typical 1", 1.5" or 2" Hamamatsu fine-mesh PMTs at various orientations 
$\theta$ (from top to bottom). Top right panel: behaviour of the gain variation 
in moderate magnetic field  for  typical 1"or 1.5" Hamamatsu fine-mesh PMTs 
at various orientations $\theta$.
Bottom panel: multiphoton timing resolution and  transit time ratio 
 as function of the magnetic field 
B for a typical 1" Hamamatsu R5505 fine-mesh PMT at various orientations 
$\theta$.
}
\label{fig-1inch}
\end{center}
\end{figure*}

Gain reduction and timing resolution were measured  
as a function of magnetic field and relative orientation angle ($\theta$)
for the three types of fine-mesh PMTs under study.
Due to the effect of magnetic field on the accelerated electrons inside the PMTs
we can expect a reduction of gain as the B field increases and also a marked
dependence of the relative gain as a function of the orientation $\theta$ angle.
The gain reduction as a function of the increasing magnetic field at
various values of $\theta$ are
 shown for typical 1",1.5",2" fine-mesh PMTs in figure
\ref{fig-1inch}~\footnote{Results for the R5924 PMTs have been reported
previously in reference \cite{belle}}.

A more detailed study of the effect of the PMT orientation with respect to the
magnetic field was performed at moderate values of the magnetic 
field (up to $\sim 500$ Gauss)
with the same 
setup used for the tests of conventional PMTs. Results for a typical
1" or 1.5" fine-mesh PMT are also reported in Figure \ref{fig-1inch}. 
The decrease of the relative gain at large values of $\theta$ is connected 
to secondary electron losses: large angle electrons fall out the grid 
surfaces and are lost. The deflection of particle trajectories is also
responsible for the enhancement of gain up to a certain value of $\theta$.
Fine-mesh PMTs are well behaving up to a critical orientation $\theta_c$
depending on the size of the projection of the photocathode onto the
anode, with respect to the magnetic field direction:
typically $\theta_c=45^0-60^0$. 
For an orientation angle $\theta$ exceeding the critical angle $\theta_c$
the gain of a fine-mesh PMT has a noticeable decrease and shows a complicate
behaviour, as the magnetic field  {\bf B} increases,
up to $\theta$ values ($\sim 80^{0}$) beyond which the gain shows a steep 
drop in gain. 
This effect may be reduced by
using a composite shielding solution similar to the one adopted 
for  conventional
PMTs. 

Timing characteristics of fine-mesh PMTs show a weak dependence from 
the field strength and direction up to $\sim 0.6-0.8$ T, in spite of the large
gain reduction, as also shown in figure \ref{fig-1inch} for $\theta \leq 
\theta_{c}$. 
\section{Conclusions}
Measurements have been performed up to longitudinal fields of 600 G 
for a sample of conventional R4998 PMTs, with different shielding options.
While a simple 1-mm thick $\mu$-metal shielding is satisfactory 
 up to 60 (150) G
for an axial (orthogonal) {\bf B} field, an additional ARMCO
($6 \times 6 \ cm^2$) shield is required for longitudinal (orthogonal) fields 
up to 600 G. 

Additional measurements for  fine-mesh PMTs in magnetic fields up to
1.2 T (see \cite{bonesini2} for additional details) show that 
fine-mesh PMTs are insensitive to longitudinal magnetic fields up to
$\sim 1$ T. However their performance degrades  quickly for orientation
angles larger than a critical value ($\theta_c$), 
even at small values of the magnetic field. 
This critical angle is typically in the range
$45-60^{0}$.

The previous results may be of interest for the optimization of time-of-flight
detectors, based on scintillation counters read by PMTs, that have to work
inside moderate magnetic field or the fringe fields of high magnetic fields. 
TOF detector timing 
resolutions show comparable results with both conventional 
R4998 PMTs or fine-mesh R5505 PMTs, see reference \cite{bertoni} for more
details. 

\section*{Acknowledgments}
 
The essential help of Dr. Y. Karadzhov (University of Sofia), Dr. M. Rayner
(University of Oxford) and Dr. G. Cecchet (INFN Pavia)  in the setting up of 
the data taking and the early stages of this work is acknowledged.
We would like to thank Dr. L. Tortora (INFN Roma Tre) who raised our attention
to the D0 note on the conventional PMT box shielding. 
We acknowledge the help of Mr. F. Chignoli and R. Mazza of INFN Milano Bicocca 
for the preparation of the mechanics setup. We are indebted
also to Dr. M. Bombonati and L. Confalonieri, Hamamatsu Italia, 
and Ing. L. Vernocchi 
for help and many enlightening discussions.

\end{document}